# Solar Still Coupled With Solar Collector and Storage Tank

[1]Rajesh .A.M , [2]Bharath .K.N
[1]Dept. of Mechanical Engineering, S.J.M.I.T, Chitradurga, Karnataka, India.
rajesh.am82@rediffmail.com
[2]Dept. of Mechanical Engineering, Univ. B.D.T College of Engineering, Davangere, Karnataka, India.
kn.bharath@gmail.com

**Abstract**

**Acute shortage of good, clean drinking water is a major problem for most developing countries of the world. In most cases, ponds, streams, wells and rivers are often polluted that they are unsafe for direct use as drinking water .Often water sources are brackish and or contain harmful bacteria. Therefore cannot be used for drinking .In addition there are many coastal locations where sea water is abundant but potable water is not available. Solar distillation is one of the important methods of utilizing solar energy for the supply of potable water to small communities where natural supply of fresh water is inadequate or of poor quality .In this direction an experimental performance analysis was carried out on a single basin still compared with FPC coupled one. Test were carried out for different water samples namely borewell water, sea water, river water for a water depth of 20 mm .Measurement of various temperatures solar intensity, distillate water collected from north and south slope were taken for several days under local climatic conditions. The study shows that single basin still productivity enhances by 42 percent for borewell water, 40 percent for sea water and 45 percent for river water when the still coupled with FPC Flat Plate Collector,The various other tests like chlorine content, Total hardness, Calcium content, Electrical conductivity, TDS, pH value, were carried out in the laboratory and found that water is safe and pure for drinking. Solar distillation becomes very attractive in expensive long term low technology system especially useful where the need for small plant exists**

**Key Words**: solar distillation, Solar still, augmentation, flat plate collector, productivity, Borewell water, River water, sea water, potable water.

## 1. INTRODUCTION

Most existing desalination plants used fossils fuels as a source of energy. The conventional distillation process namely reverse osmosis, electro dialysis, multieffect evaporation etc are not only energy intensive but also uneconomical when the demand for the fresh water is small [1].Solar distillation is the only attractive process for saline/brackish water by using solar energy .The basin time solar stills are simple in design, manufacturing, operation and economical .But the productivity of fresh water is low on an average of 2.5 $l/m^2$ day. Enhancing the still's yield has been studied by several investigators, suggesting various approaches. M.S. Sodha et al [2] analysized solar still with double roof .R.A Collins et al [3] made tests on a simple solar still coupled with an external condenser. M .Boukar et al [4] made comparative study on simple basin solar still. In the present work an experimental study was carried out to compare the performance of single basin double slope solar still and FPC coupled one.

## 2. METHODOLOGY

The experimental set up is one kind of a so called active distillation system where a conventional solar still is assisted by a heat source- flat plate collector.

The set up fabricated local and is shown in fig.1 The solar still is covered with double glass in inverted V- type at $35^o$ slope. The basin area is of 1 $m^2$ .It is painted with black paint and insulated with rock wool. The collector, area 1.15 $m^2$ made of MS sheet of 0.5 mm thick, is coated with black paint .Eight copper tubes of 15mm diameter are fixed to the collector plate with pitching o tubes 12 cm.

The collector is placed facing due south at tilt angle of $22^o$ (fixed).The solar still was positioned along east – west direction with glass cover facing north – south direction to have maximum yield per day. The distilled was collected in two separate graduated cylinders from both north and south sides.

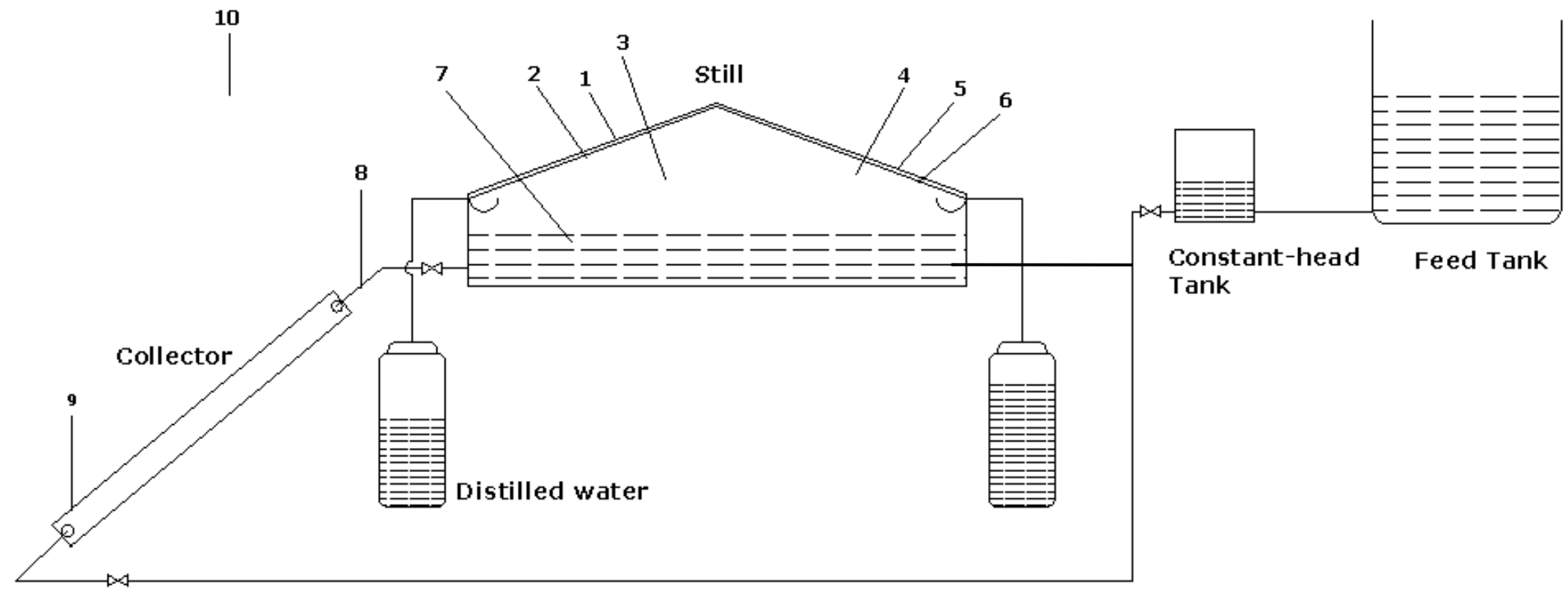

Figure 1a. Schematic diagram of experimental set up showing the location of thermocouples [with collector] (1-10 are thermocouple locations)

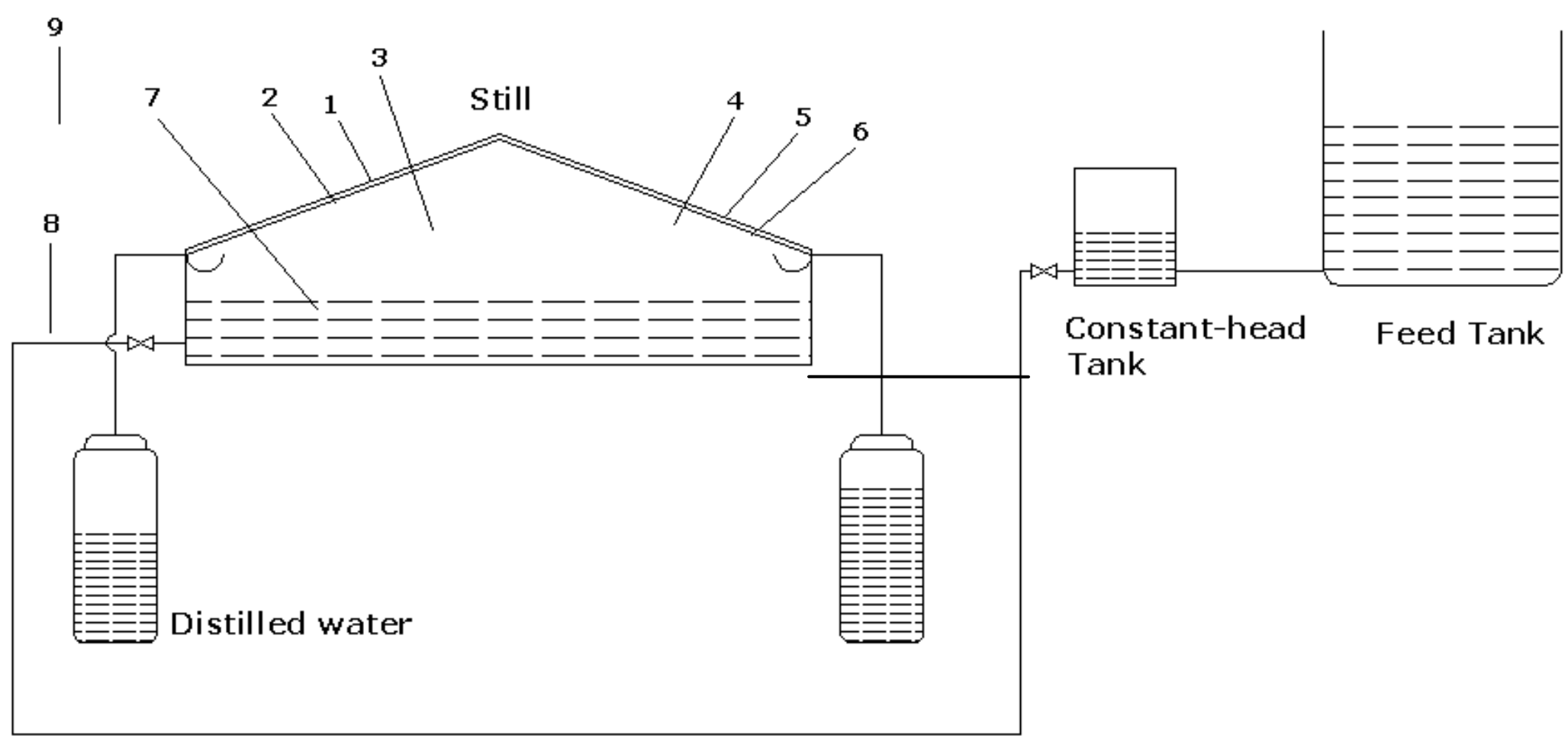

Figure 1b. Schematic diagram of experimental set up showing the location of thermocouples [solar still alone] (1-8 are thermocouple locations)

The entire system was made leak proof .Thermocouples 10 numbers were fixed at various points to measure basin water temperature, evaporative basin chamber temperature(north and south),inside and outside glass cover temperature(north & south),flat plate collector inlet & outlet temperature,F.P.C. glass outside temperature, ambient temperature. The systems were operated with and without coupling FPC at location Davanagere (14.31$^{o}$N, 75.58$^{o}$E) under local conditions for a water depth of 20mm for different water samples namely bore well, river water. Sea water.

## 3. RESULTS AND DISCUSSIONS

Experiments were conducted for both the cases of solar still alone and FPC coupled one for both bore water and river water for a water depth of 20mm. For borewell water, the various temperatures of basin water, inside and outside glass cover vapour and ambient versus time for both the cases are shown in fig.2(a), (b) Shows the variation of hourly temperature for hard water with coupled with FPC and solar still alone respectively and fig.2(c),(d) shows for river water and fig.2(e),(f) shows for sea water .Similar trends were noticed for all other experiments .The basin water temperature was the highest among all the temperatures noticed and always occurred between 13 to 14 hrs.For borewell water it ranged between 62$^{o}$c to 65$^{o}$c for still alone and 66$^{o}$c to 72$^{o}$c for FPC coupled one. For river water it ranged between 63$^{o}$ to 65$^{o}$c for still alone and 68$^{o}$c to 70$^{o}$c for FPC coupled one .for sea water it ranged between 60$^{0}$ to 62$^0$c for still alone and 65$^o$c to 68$^o$c for FPC coupled one and Ambient temperature was in the range of 29$^o$c to 33$^o$c .the basin water temperature for sea water was low compare to other two water and it was noticed that as salt content in the water increases the productivity was less The amount of distilled water collected from the north side of the still was more compared to south side in both cases as shown in fig.3 (a), (b) , fig.4 (a) (b). fig.5 (a) (b). The Rate of evaporation, Amount of energy utilized in vaporizing the water are tabulated in the table.1

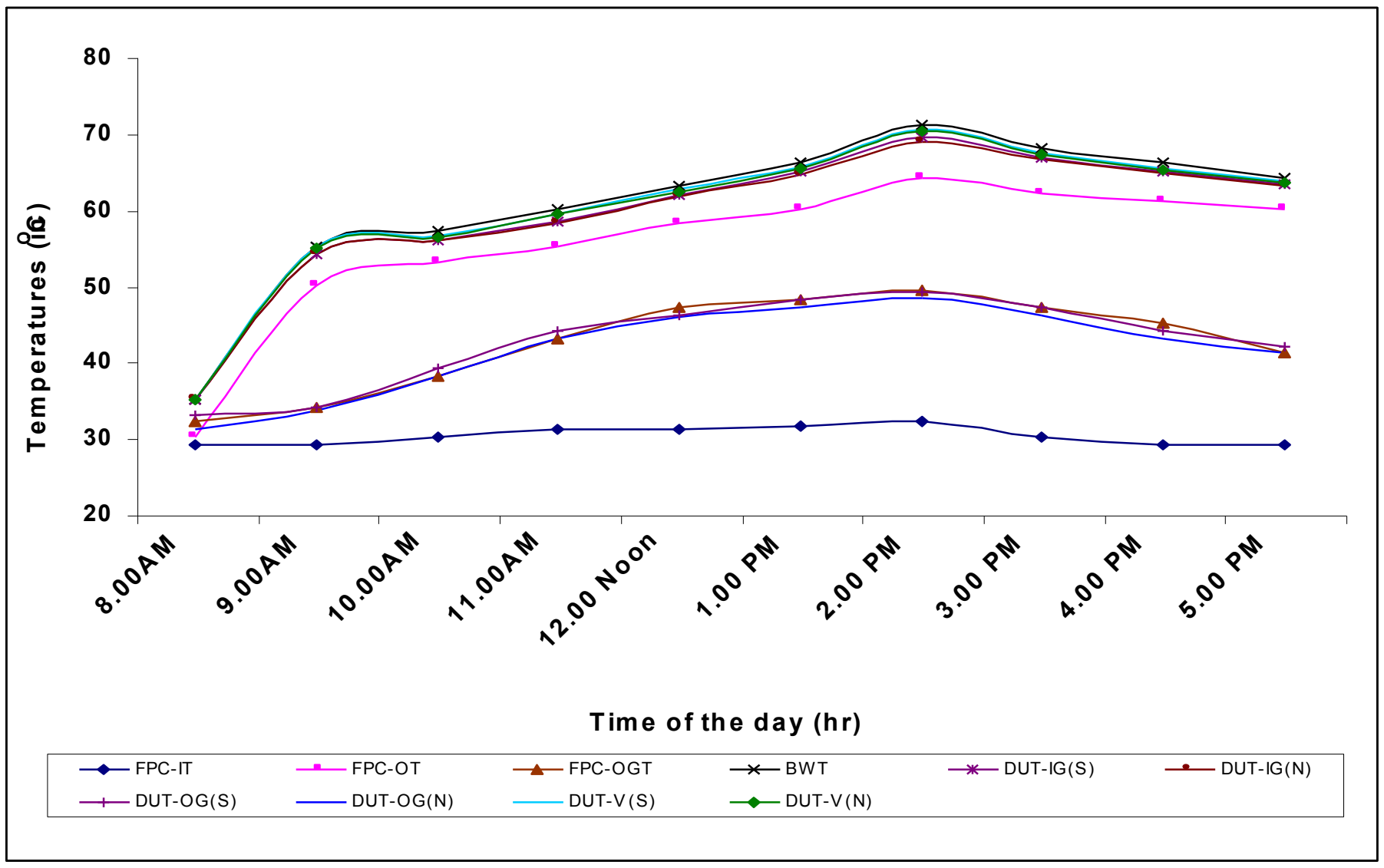

Figure 2(a). The variation of hourly temperature for hard water coupled with FPC

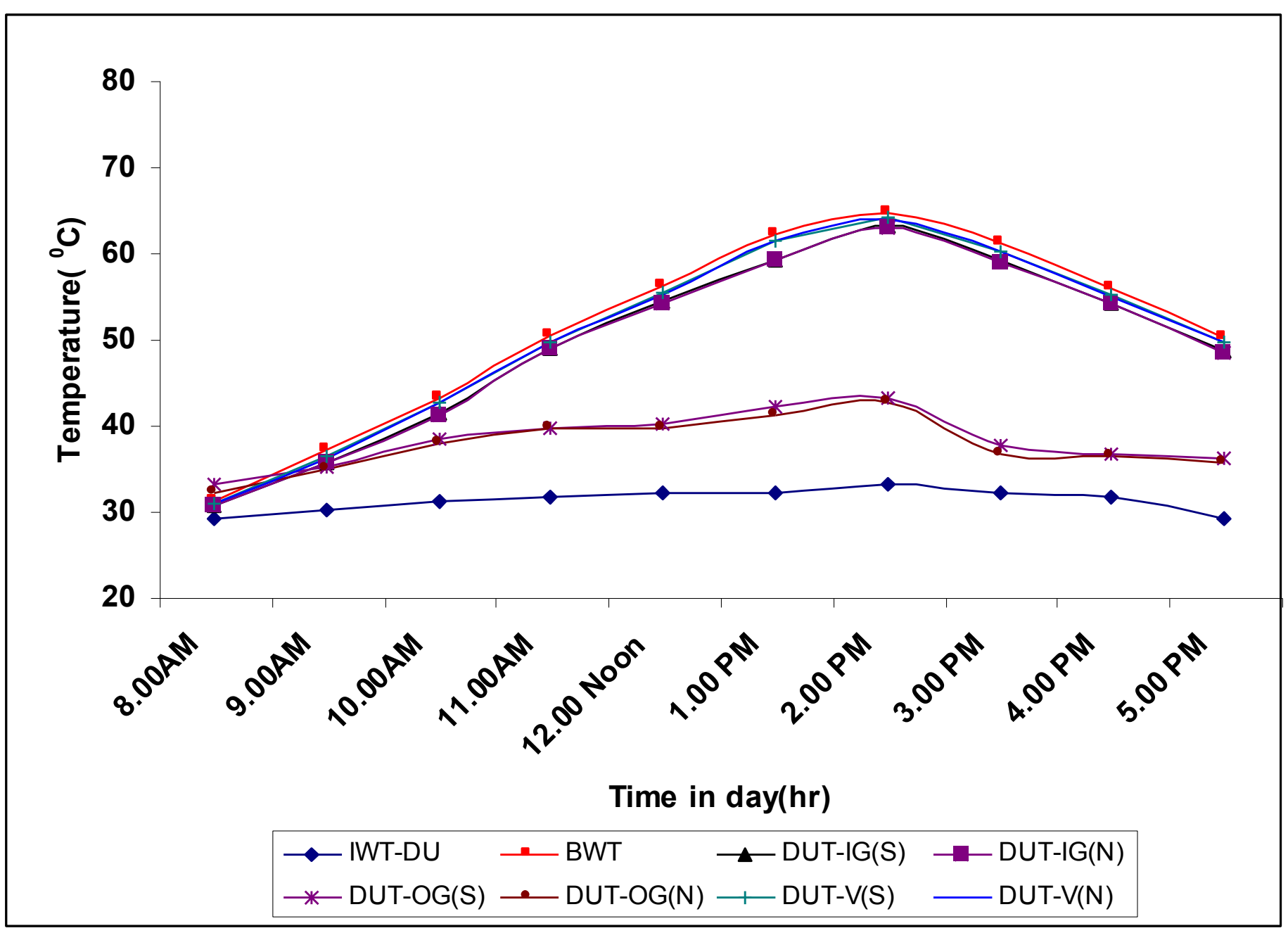

Figure 2(b). The variation of hourly temperature for hard water with solar still alone

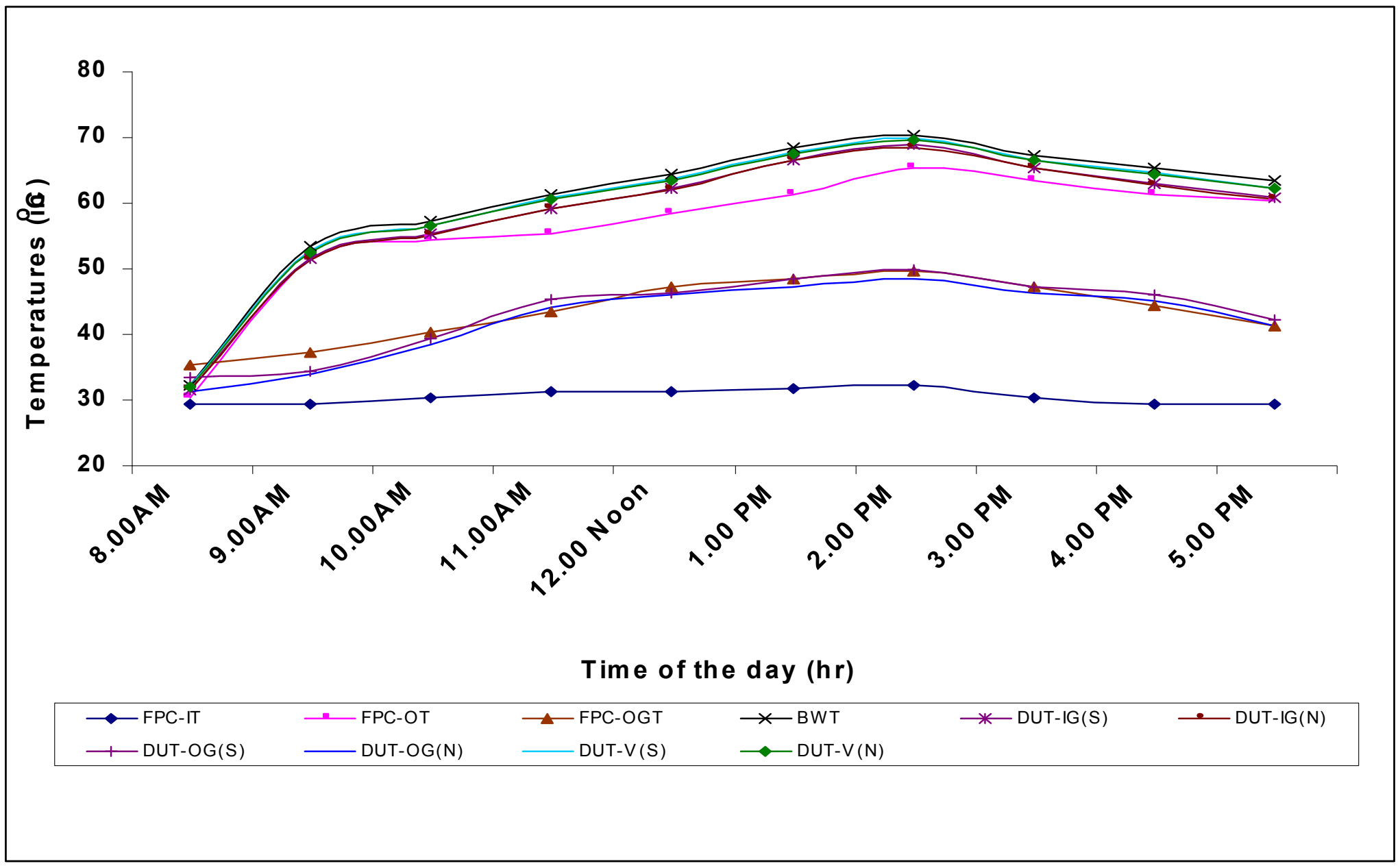

Figure 2 (c). The variation of hourly temperature for river water coupled with FPC

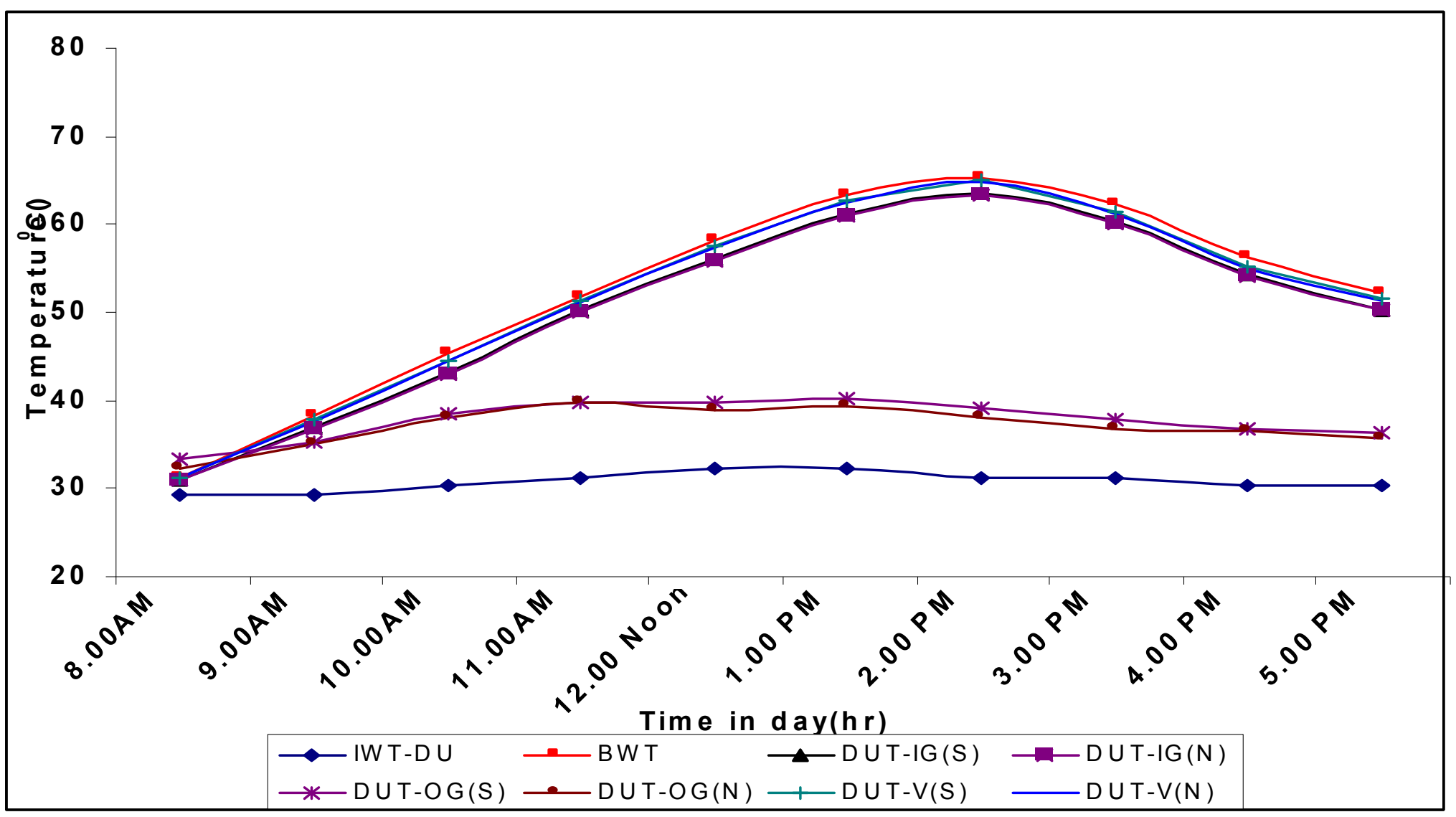

Figure 2 (d). The variation of hourly temperature for river water with solar still alone

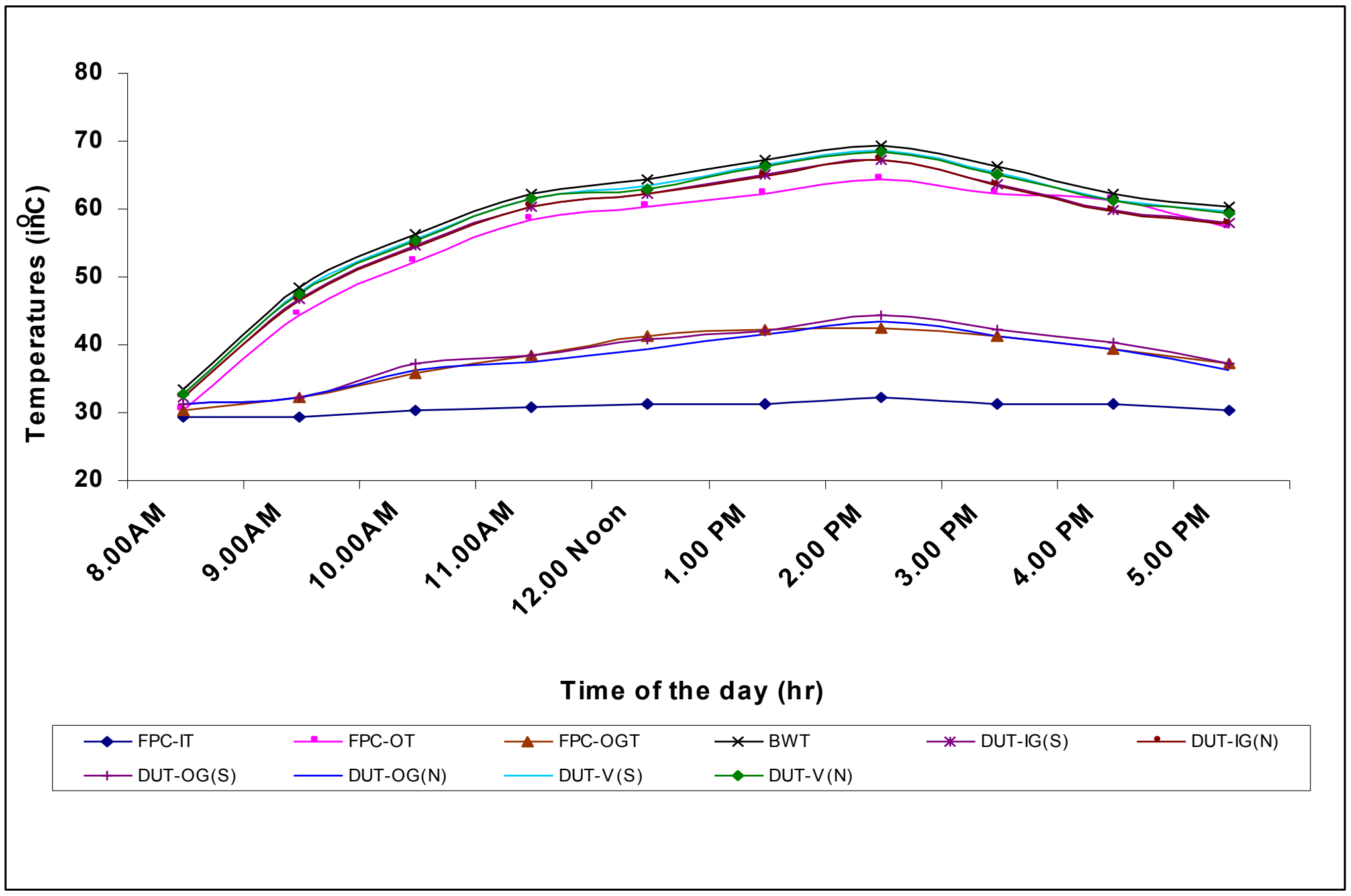

Figure 2 (e). The variation of hourly temperature for sea water coupled with FPC

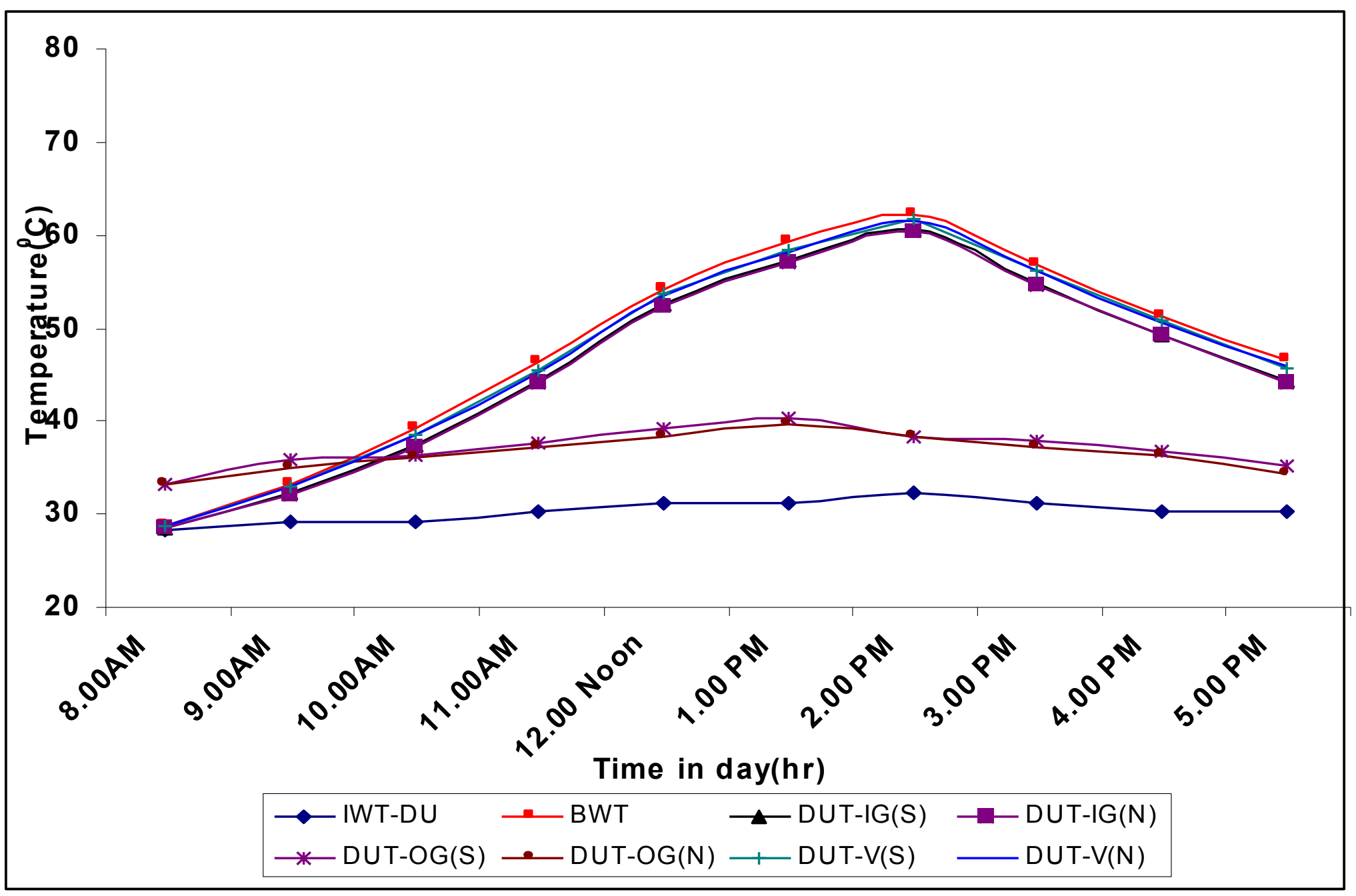

Figure 2 (f). The variation of hourly temperature for sea water with solar still alone

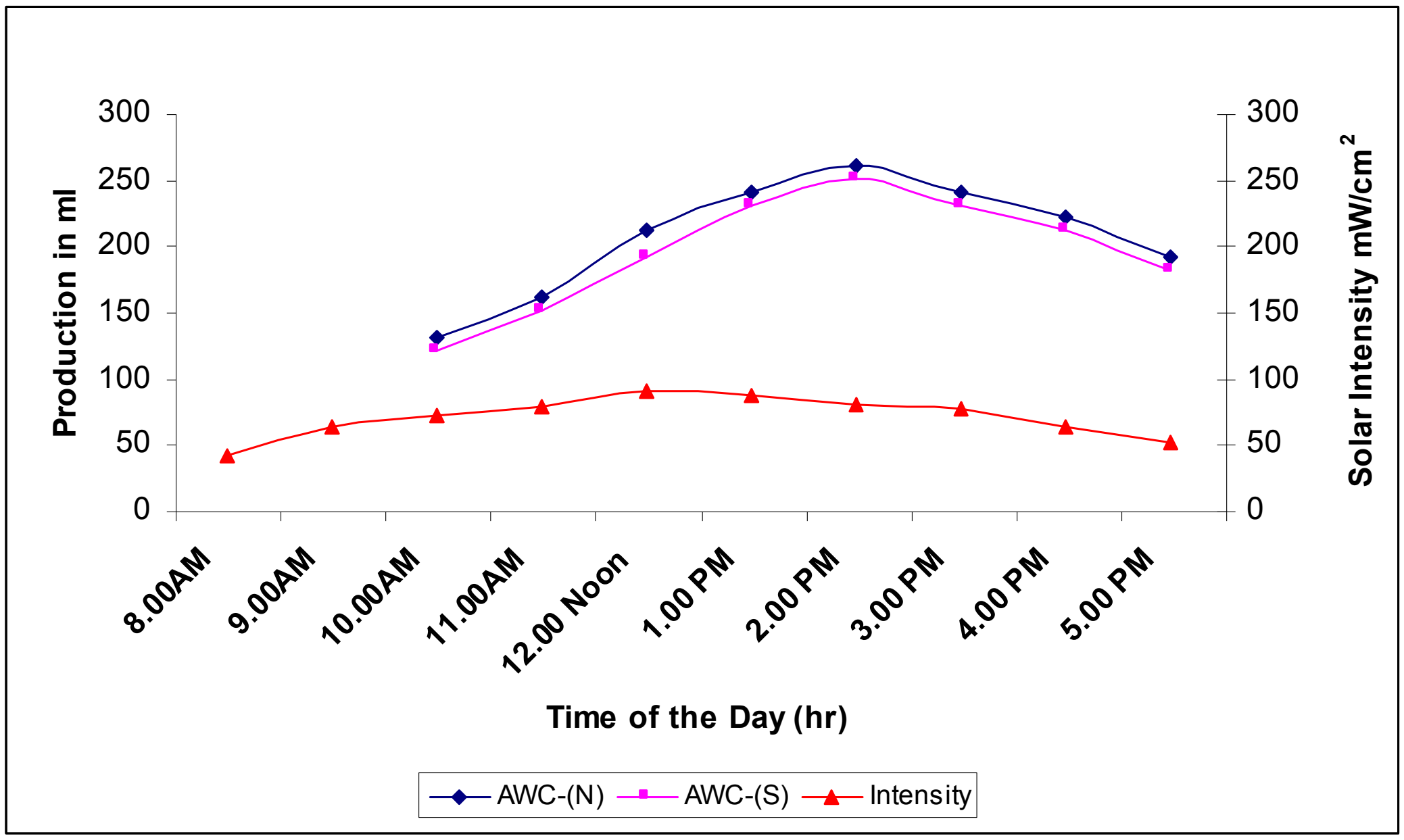

Figure 3 (a). The hourly variation of various parameters coupled with FPC [hard water]

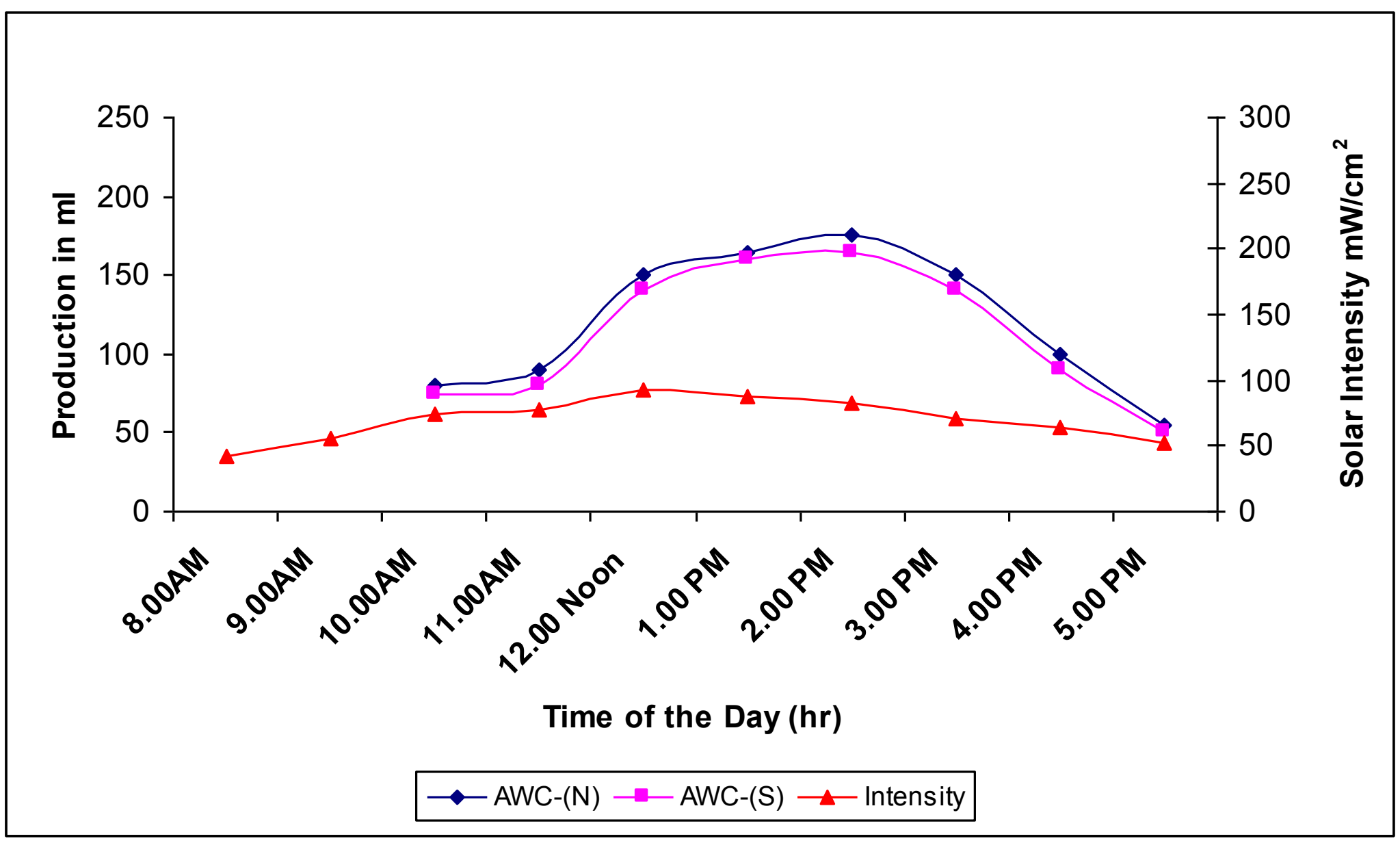

Figure 3 (b). The hourly variation of various parameters with solar still alone [hard water]

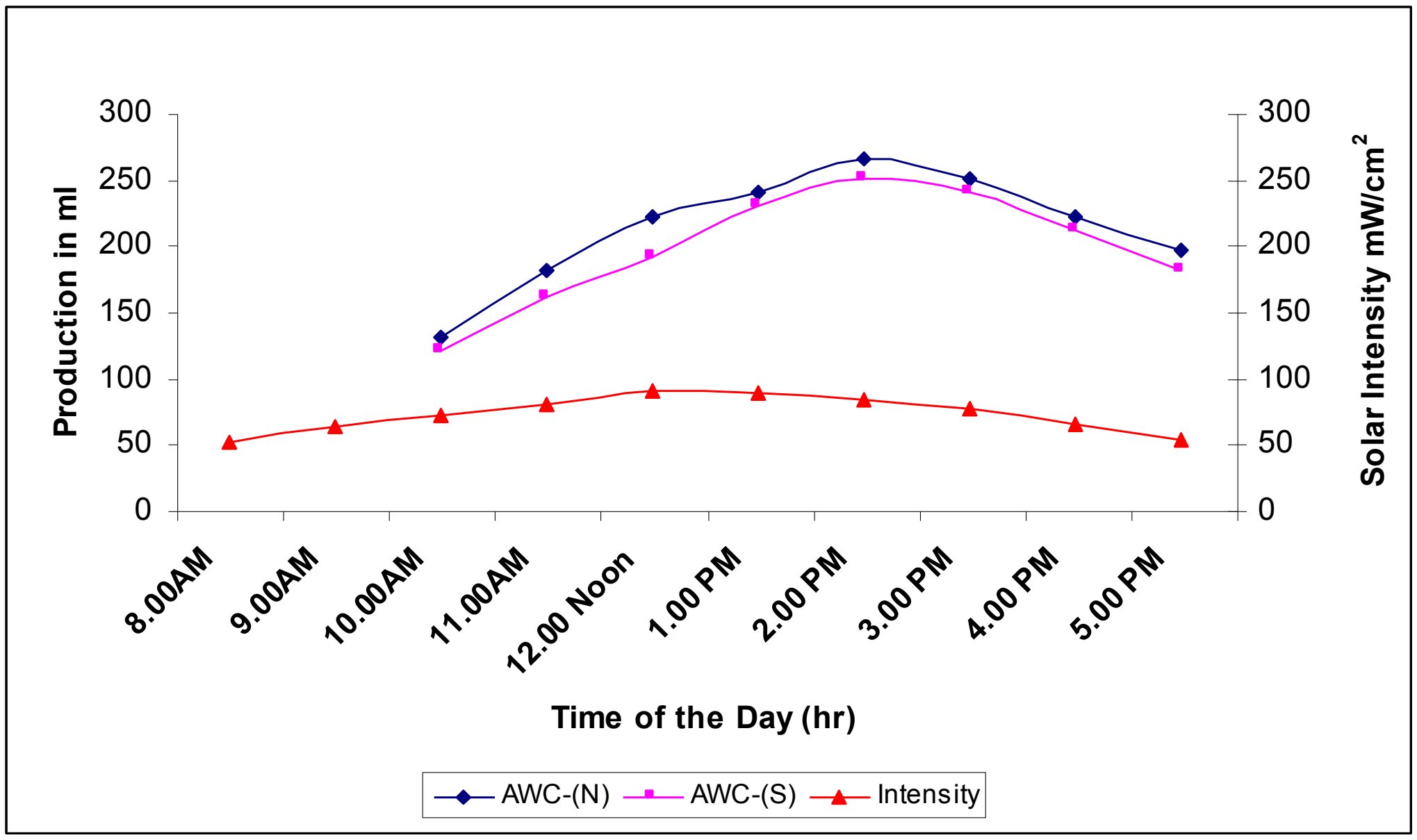

Figure 4 (a). The hourly variation of various parameters with hybrid unit [river water]

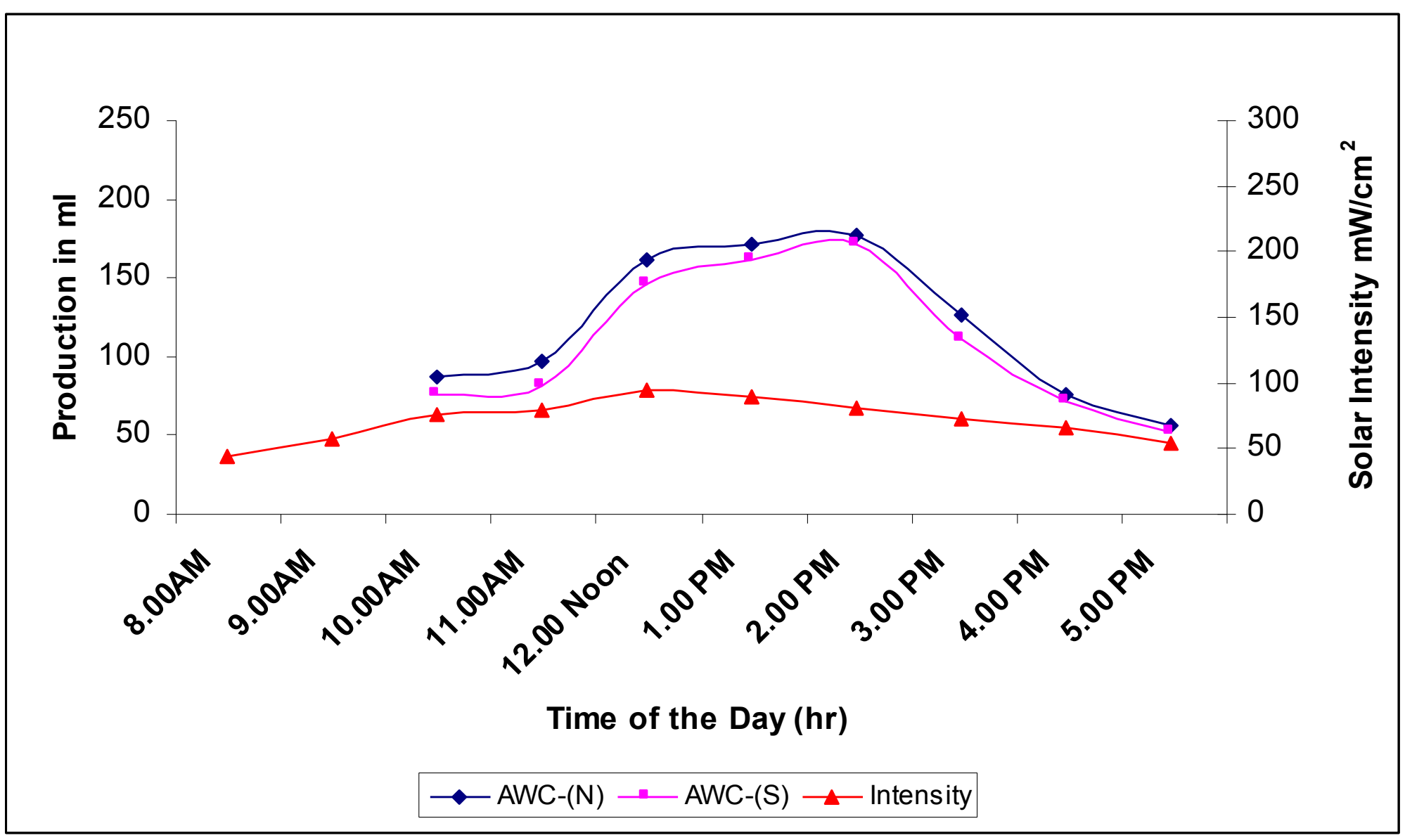

Figure 4 (b). The hourly variation of various parameters with solar still alone [river water]

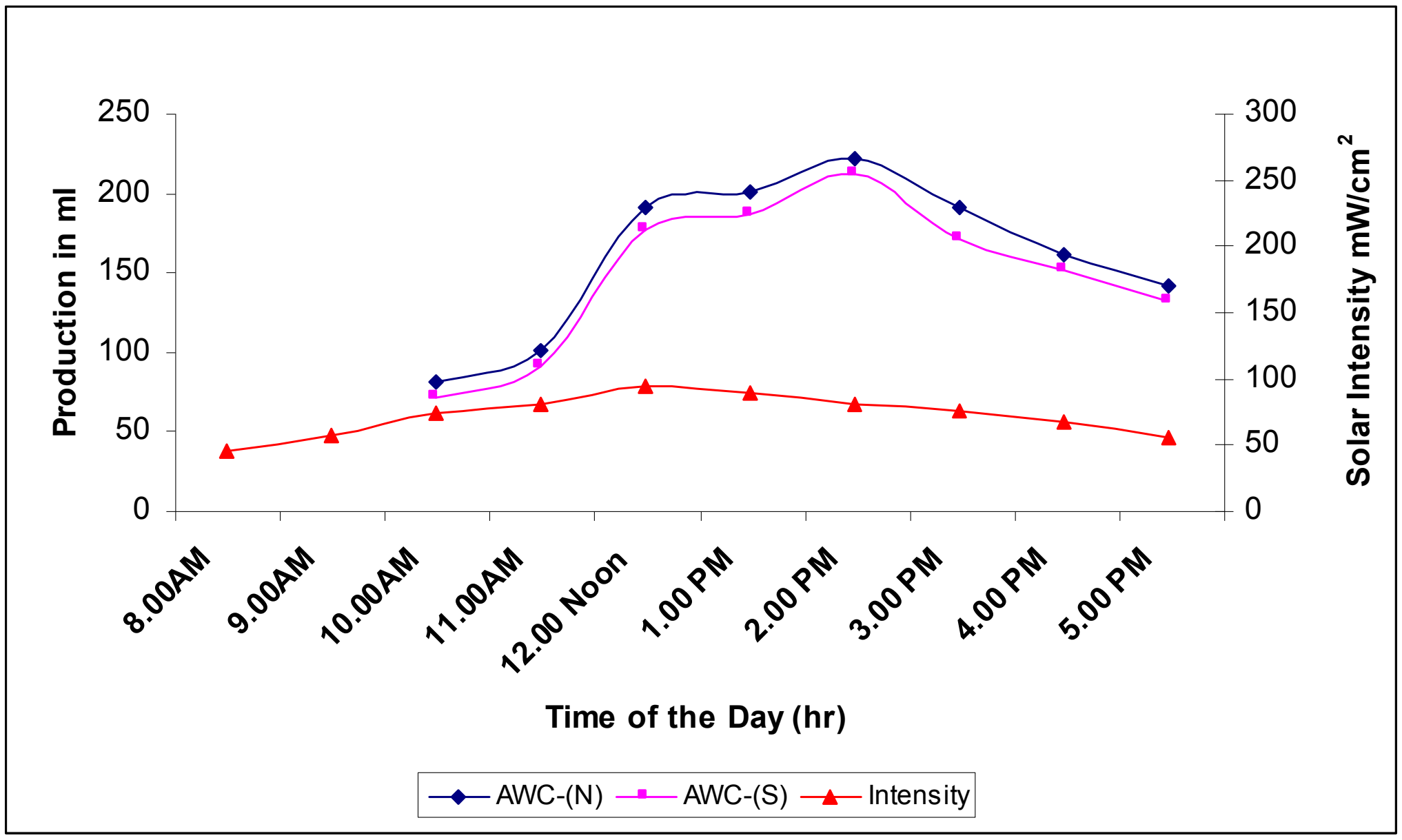

Figure 5 (a). The hourly variation of various parameters with hybrid unit [sea water]

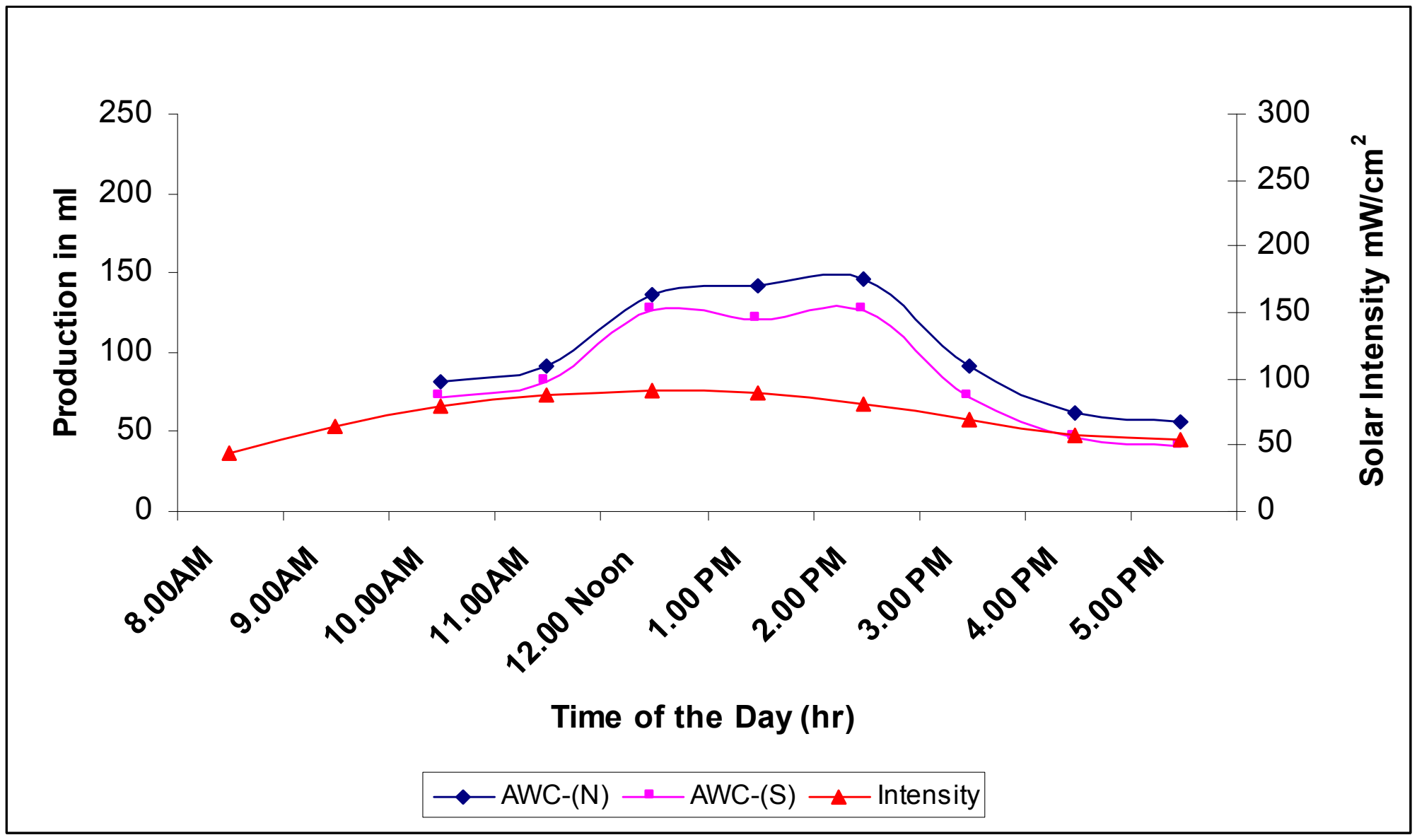

Figure 5 (b). The hourly variation of various parameters with solar still alone [sea water]

Table 1. Rate of evaporation, Amount of energy utilized in vaporizing the water are tabulated

| Type of water and operation | Production rate, in liters | Rate of Evaporation In liters per hour | Qe=amount of energy utilized in vaporizing the water ($KW/m^2$), | i Percentage of Enhancement in the mass of distilled water using hybrid unit |
|---|---|---|---|---|
| River water with hybrid unit | 3.280 | 0.36444 | 8.5574 | 45.12 |
| River water with still alone | 1.800 | 0.2 | 4.1553 | |
| hard water with FPC | 3.210 | 0.3566 | 8.375 | 43.01 |
| Hard water with still alone | 1.765 | 0.1961 | 4.098 | |
| Sea water with hybrid unit | 2.460 | 0.27333 | 6.418 | 40 |
| Sea water with still alone | 1.470 | 0.1633 | 3.835 | |

Table 2: The acceptable characteristics for potable/edible water as per BIS is given

| **Sl No** | **Characteristics** | **Acceptable** | **causes for Rejection** | **Remarks** |
|---|---|---|---|---|
| 1 | Turbidity (NTU) | 5 | 10 | Consumer acceptance decrease |
| 2 | Colour | 5 | 25 | |
| 3 | PH | 6.6 to 8.5 | No relaxation | Water will affect the mucous membrane |
| 4 | Total dissolved solids(mg/l) | 500 | 2000 | Consumer acceptance decreases. Laxative effect upon people who are not accustomed to it. Mau cause gastro intestinal irritation |
| 5 | Total hardness (as $CaCo_3$) (mg/1) | 300 | 600 | Encrustations in water supply structure and adverse effect on domestic use/scale formation |
| 6 | Chlorides (as Cl)(mg/1) | 250 | 1000 | Taste, palpability and corrosion are effected |
| 7 | Fluorides (as F)(mg/1) | 1.0 | 1.5 | Results in dental/sketal fluorisis |
| 8 | Calcium (as Ca) (mg/1) | 75 | 200 | Encrustation in water supply structure and adverse effects on domestic use |
| 9 | Magnesium (as Mg) (mg/1) | <=30 | 100 | |

Table 3 water analysis for non treated water

| Sl. No | Characteristics | River Water | Borewell water | Sea water |
|---|---|---|---|---|
| 1 | Electrical Conductivity at 30 °C in(us) | 649uS | 876µS | 17mS |
| 2 | Total Dissolved Solids at 30°C in(ppm) | 70ppm | 490 ppm | 9.11 PPT[part per tonne] |
| 3 | Total hardness content in mg/l(ppm) | 48 | 408.00 | 4200 |
| 4 | Chloride content in mg/l(ppm) | 109 | 137 | 8936 |
| 5 | Calcium content in mg/l(ppm) | 4.8 | 72 | 400 |
| 6 | Magnesium content in mg/l(ppm) | 11.52 | 97.92 | 1008 |
| 7 | Ph value | 7 | 7.6 | 8.7 |

Table 4 water analysis for distilled water

| Sl. No | Characteristics | River Water | Borewell water | Sea water |
|---|---|---|---|---|
| 1 | Electrical Conductivity at 30 °C in(us) | 32.20uS | 34.80µS | 81µS |
| 2 | Total Dissolved Solids at 30°C in(ppm) | 28ppm | 55 ppm | 45 ppm |
| 3 | Total hardness content in mg/l(ppm) | 12 | 14 | 36 |
| 4 | Chloride content in mg/l(ppm) | 12 | 14 | 16 |
| 5 | Calcium content in mg/l(ppm) | 3.2 | 4.8 | 5.6 |
| 6 | Magnesium content in mg/l(ppm) | 2.88 | 3.36 | 8.64 |
| 7 | Ph value | 6.0 | 6.0 | 7 |

## WATER ANALYSIS

**Water Analysis Test Report [pre treatment of raw water and post treatment of distilled water]:**

To determine the purity of distilled water collected from the solar distillation unit. The following tests have been carried for the water entering and leaving the distillation unit Chloride test, Total hardness test, Calcium in water, TDS [total dissolved solids], Magnesium, Electrical conductivity, pH.

## 5. CONCLUSIONS

1. From the above results and discussions it is cleared that production rate was higher on north side compared to south side.
2. The production rate varies accordingly with respect to solar intensity. The productivity of the still increases by 42 % in case of bore well water , 45% in case of river water and 40% in case of sea water when FPC .
3. This is due to enhancement of basin water temperature by FPC .Amount water collected primarily depends on the peak temperature and length of duration of highest temperature of basin water.
4. The hybrid solar still design tested in this study confirms production of high quality drinking water from source water of very poor quality.
5. The solar distillation plants are relatively inexpensive low temperature technology system, and would be one of the best solutions to supply fresh drinking water to small isolated communities with no technical facilities

## BIOGRAPHY

**Rajesh.A.M** , completed his M.Tech in Thermal Engg. in U.B.D.T.C.E, Davangere, India and presently working as a Lecturer in dept. of mechanical Engg. in S.J.M Institute of Technology, Chitradurga, Karnataka, India.

**Bharath.K.N,** completed his M.Tech in machine design in U.B.D.T.C.E, Davangere, India and presently working as a Lecturer in dept. of mechanical Engg. in University B.D.T College of Engineering, Davangere, Karnataka, India.